\documentclass[a4paper,11pt]{article}
\usepackage{pos}

\title{Non-relativistic QCD Study of Excited Bottomonia at Finite Temperatures on a Fine Lattice}
\ShortTitle{NRQCD Study of Excited Bottomonia}

\author[1]{Heng-Tong Ding}
\author*[1]{Wei-Ping Huang}
\author[2]{Rasmus Larsen}
\author[3]{Stefan Meinel}
\author[4]{Swagato Mukherjee}
\author[4]{Peter Petreczky}

\affiliation[1]{Key Laboratory of Quark and Lepton Physics (MOE) and Institute of
    Particle Physics, \\
    Central China Normal University, Wuhan 430079, China}    
\affiliation[2]{Fakult\"at f\"ur Physik, Universit\"at Bielefeld, D-33615 Bielefeld, Germany}
\affiliation[3]{Department of Physics, University of Arizona, Tucson, Arizona 85721, USA}
\affiliation[4]{Physics Department, Brookhaven National Laboratory, Upton, New York 11973, USA}

\emailAdd{huangweiping@mails.ccnu.edu.cn}

\abstract{The temperature dependence of bottomonium correlators up to the 3S and 3P excited states are presented in the range $T \simeq 133-250$ MeV. These lattice calculations employ the non-relativistic QCD (NRQCD) approach for bottom quarks on (2+1)-flavor gauge backgrounds, using the highly improved staggered quark (HISQ) action near the physical point. The study utilizes a fine lattice spacing of 0.0493 fm at all temperatures. Extended bottomonium operators are implemented to achieve optimized overlaps with the targeted excited states, enhancing sensitivity to thermal effects. To probe in-medium modifications of excited bottomonia, we extract thermal widths and in-medium masses from bottomonium correlators, parameterizing the spectral function with a Gaussian ansatz. Our results confirm nonzero thermal widths for various bottomonium states as the temperature increases, while no significant mass shifts are observed. Additionally, we check that the in-medium properties of bottomonia are almost not affected by variations in the choice of extended operators.}

\FullConference{The 41st International Symposium on Lattice Field Theory (LATTICE2024)\\
 28 July - 3 August 2024\\
Liverpool, UK\\}

\begin{document}
\maketitle

\section{Introduction \label{sec:intro}}
Quarkonium suppression, proposed as strong evidence for the existence of quark-gluon plasma (QGP) produced in high-energy nuclear collisions~\cite{Matsui:1986dk,Satz:2006kba}, has been experimentally observed~\cite{Arnaldi:2009ph,PHENIX:2008jgc,CMS:2011all,CMS:2012gvv,Wang:2019vau}. This proposal is due to a widely accepted picture that color screening within the deconfined QGP medium will make the range of quark-antiquark force shorter than the characteristic size of a quarkonium state, finally leading to its dissolution. Then, a further consequence is that a loosely-bound quarkonium state, such as the excited states, is more susceptible to in-medium modification than a compact one with a smaller size like the ground states. This leads to the expectation of sequential in-medium modification of different quarkonium states based on their size hierarchy~\cite{Karsch:2005nk}, supported by measurements at CMS~\cite{CMS:2011all,CMS:2012gvv} and STAR~\cite{Wang:2019vau}. Given these considerations, probing the in-medium properties of quarkonia with various sizes characterizes QGP on different length scales and has brought together extensive experimental and theoretical efforts; see, e.g., Refs.~\cite{Aarts:2016hap,Mocsy:2013syh} for reviews.

Theoretically, although the consensus that quarkonium will eventually dissolve in the QGP at sufficiently high temperature can be reached, the microscopic mechanism behind the in-medium modification of quarkonia is more complicated than the original picture proposed in Ref.~\cite{Matsui:1986dk} and still inconclusive. Recent works reveal that at nonzero temperatures, the quark-antiquark potential becomes complex, acquiring a large imaginary component that encodes the dissipative effects of the medium, without indications of color screening in its real part modification~\cite{Laine:2006ns,Brambilla:2008cx,Bala:2021fkm,Bazavov:2023dci}. Thus to gain more information about the modification of quarkonia in the medium, detailed investigation of their properties at finite temperatures is needed.

In-medium properties of quarkonia are encoded in the spectral function, $\rho(\omega, T)$, which is related to the corresponding Euclidean correlator, $C(\tau, T)$, through the following Laplace transform:
\begin{align}
    C(\tau, T)=\int_{0}^{\infty} \mathrm{d} \omega ~ \frac{\cosh[\omega(\tau-1/2T)]}{\sinh(\omega/2T)}  \rho(\omega, T) \,.
    \label{eq:definition-relativistic-spectra-function}
\end{align}
Here $C(\tau, T)$ can be measured through lattice-regularized field theory simulations. However, the calculation of bottomonium correlators on the lattice is challenging. The relativistic lattice treatment of quarkonia with heavy quark mass $m_q$ will lead to significant systematic discretization effects proportional to $am_q$~\cite{Jakovac:2006sf,Petreczky:2010yn}, with $a$ the lattice spacing. Besides, extracting in-medium properties from a finite number of discrete lattice data points for $C(\tau, T)$ is also difficult, due to the limited sensitivity of $C(\tau, T)$ to the thermal effect~\cite{Mocsy:2007yj,Petreczky:2008px}. This stems from the fact that the maximal temporal extent $\tau_{\rm max}$ of $C(\tau, T)$ equals half the inverse temperature, $1/2T$ (c.f.,~\autoref{eq:definition-relativistic-spectra-function}), resulting in a shrinking temporal range of correlators as the temperature increases. 

Owing to the large mass $m_q$ and relatively small heavy-quark velocity $v$ inside quarkonium, physics at different energy scales can be separated following the hierarchy $m_q \gg m_qv \gg m_qv^2$. An effective field theory, non-relativistic QCD (NRQCD), can be reached by integrating out the ultraviolet degrees of freedom related to $m_q$~\cite{Lepage:1992tx,Thacker:1990bm}. Simulating the heavy-quark bound states through the NRQCD approach on the lattice will not only enable well-controlled discretization errors, but also make the correlators more sensitive to the thermal modifications compared to a relativistic treatment~\cite{Aarts:2010ek,Aarts:2014cda,Kim:2014iga,Kim:2018yhk}. This arises from the forbiddance of heavy-quark pair creation in such a non-relativistic formalism, leading to the periodic boundary condition in time not to be satisfied for the heavy-quark fields~\cite{Brambilla:2008cx} and \autoref{eq:definition-relativistic-spectra-function} to become~\cite{Aarts:2010ek}
\begin{align}
	C(\tau, T)=\int_{-\infty}^{+\infty} \mathrm{d} \omega~ \rho(\omega, T) e^{-\tau \omega} \,.
    \label{eq:definition-nonrelativistic-spectra-function}
\end{align}
As a result, the maximal temporal extent will be $1/T$, twice larger than that in the relativistic case.

During the calculations of quarkonium correlators, the commonly used point-like operators do not have good overlap with targeted meson states with particular quantum numbers, especially for the excited states. Thus, correlators at large $\tau$ are needed to eliminate contamination from high states and the continuum. Recent lattice NRQCD studies using extended meson operators~\cite{Larsen:2019bwy,Larsen:2019zqv} show optimal projection onto different S- and P-wave bottomonia and more sensitivity to thermal effects is obtained. Consequently, in the work we will measure the bottomonium correlators through the NRQCD approach on the lattice with extended meson operators, to gain more sensitivity to the in-medium modifications, with the aim to extract in-medium properties of bottomonia at finite temperatures. 

This proceedings contribution is based on Ref.~\cite{Ding:2025fvo} and is organized as follows. In \autoref{sec:setup} we will show detailed information about the setup of the lattice NRQCD simulations, as well as the extended meson operators used in this work. \autoref{sec:result} will present the bottomonium correlators in vacuum and at finite temperatures and the temperature dependence of in-medium parameters extracted from the continuum-subtracted correlators. Conclusions will be given in \autoref{sec:conclusion}.

\section{Details of lattice NRQCD simulations \label{sec:setup}}
To calculate bottomonium correlators on the lattice, for the bottom-quark part we used the tree-level tadpole-improved NRQCD action including spin-dependent $v^6$ corrections, as used in Refs.~\cite{Larsen:2019bwy,Larsen:2019zqv}. These calculations were performed on background gauge fields including (2+1)-flavors of dynamical sea quarks, implemented using the highly improved staggered quark and the tree-level Symanzik gauge (HISQ/tree) action. The strange-quark mass $m_s$ was fixed to its physical value with light-quark masses equal to $m_s/20$, corresponding to the Goldstone pion mass of approximately 160 MeV, close to the physical point. In this work, the spatial extent of the lattices was fixed to $N_{\sigma}=64$ and the temperature was varied by varying the temporal extent $N_{\tau}$ from 30 to 16 at a fixed lattice spacing $a=0.0493$ fm, corresponding to temperature ranging from 133 to 250 MeV.

Two types of extended meson operators proposed in Refs.~\cite{Larsen:2019bwy,Larsen:2019zqv} were used in this work. One is a Gaussian-smeared source~\cite{Larsen:2019bwy} optimized for the ground S- and P-wave bottomonium states, expressed as $O_{\alpha}(\mathbf{x},\tau)=\tilde{\bar{q}}(\mathbf{x},\tau)\Gamma_{\alpha}\tilde{q}(\mathbf{x},\tau)$, where $\Gamma_{\alpha}$ denotes different bottomonium interpolators and $\tilde{q}$ and $\tilde{\bar q}$ stand for smeared quark and anti-quark fields. The smeared quark field is formulated as $\tilde{q}=W q$~\cite{Larsen:2019bwy}, where $W=[1+ \sigma^2 \Delta^{(2)} / (4N)]^N$ with $\Delta^{(2)}$ being the covariant lattice Laplacian. For large enough $N$, this produces a smeared quark field with Gaussian profile having a width $\sigma$. We adopt the same smearing parameters $\sigma$ and $N$ to construct $W$ as those specified in Ref.~\cite{Larsen:2019bwy}. 

The other type of operator is the wave-function-optimized source~\cite{Larsen:2019zqv} given by $O_{\alpha,i}(\mathbf{x}, \tau)=\sum_{\mathbf{r}} \Psi_{i}(\mathbf{r}) \bar{q}(\mathbf{x}+\mathbf{r}, \tau) \Gamma_{\alpha} q(\mathbf{x}, \tau)$, where $i$ represents different excited states for a specific bottomonium channel $\alpha$, with $i$ ranging from 1S(P) to 3S(P) states in this work. The wave function $ \Psi_{i}(\mathbf{r})$, acting as a shape function, is solved from the discretized three-dimensional Schr\"odinger equation, where an $\mathcal{O}(a^4)$-improved discretized Laplacian is used with the same spacing as the gauge backgrounds and the potential takes a Cornell form with the same parameterization as in Refs.~\cite{Meinel:2010pv,Larsen:2019zqv}. Measurements with such optimized source will result in nonzero off-diagonal correlators $C_{\alpha,ij}(\tau) = \Sigma_{\mathbf{x}} \langle O_{\alpha,i}(\mathbf{x}, \tau) O^{\dagger}_{\alpha,j}(\mathbf{0}, 0) \rangle$ for $i\neq j$. Thus, variational analysis is used to obtain the diagonalized correlator $C_{\alpha,i}(\tau)=\sum_{\mathbf{x}}\langle \Omega_{\alpha,ij} O_{\alpha,j}(\mathbf{x},\tau) O^{\dagger}_{\alpha,k}(\mathbf{0},0)\Omega^{\dagger}_{\alpha,ki}\rangle$, with $\Omega_{\alpha,ij}$ solved from the generalized eigenvalue problem~\cite{Larsen:2019zqv,Blossier:2009kd,Orginos:2015tha,Nochi:2016wqg}. For a targeted bottomonium state $\alpha$, the rotation matrix $\Omega_{\alpha,ij}$ is computed at zero temperature and uniformly applied across all temperatures. 

For the above two types of extended sources, bottomonium correlators were measured using 24 sources distributed across different locations for each gauge configuration to improve the signal. The bottom-quark mass, $M_b$, as a free parameter in the NRQCD action, is fixed by matching the kinetic mass of the $\eta_b$ to its PDG value~\cite{ParticleDataGroup:2024cfk}, leading to $aM_b=0.955(17)$ for spacing $a=0.0493$ fm, which is consistent with previous value $aM_b=0.957(9)$~\cite{Larsen:2019bwy}. Thus, $aM_b=0.957$ was adopted in this work. More details about the lattice NRQCD calculations can be found in Ref.~\cite{Ding:2025fvo}.

\section{Results \label{sec:result}}

\subsection{Vacuum case \label{subsec:vacuum}}
\begin{figure*}[!htp]
  \centering
    \includegraphics[width=0.45\textwidth]{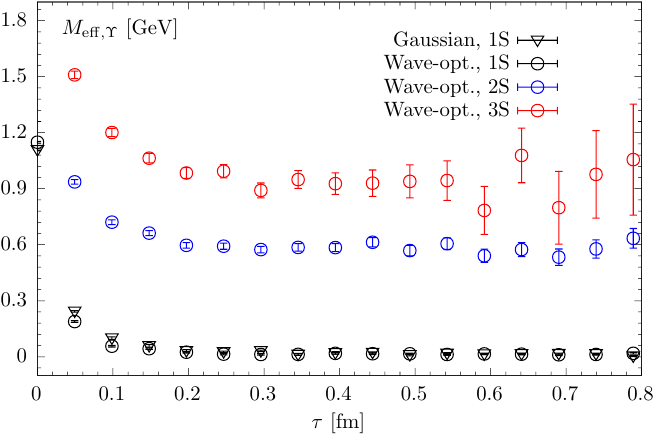}
    \includegraphics[width=0.45\textwidth]{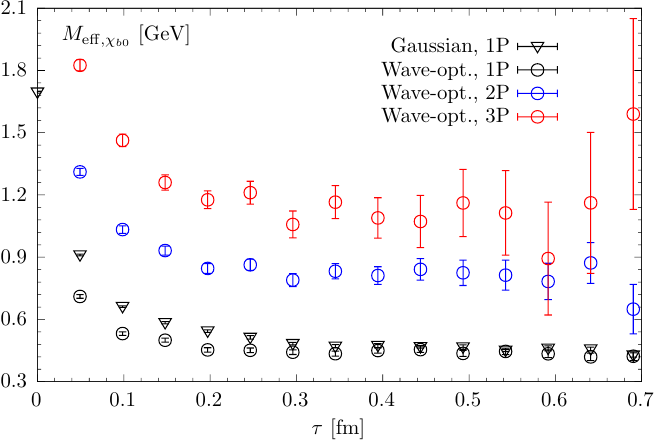}
  	\caption{The effective masses at zero temperature for $\Upsilon$ (left) and $\chi_{b0}$ (right) correlators, from Gaussian-smeared extended sources (triangle) and wave-function-optimized sources (circle) respectively. The vertical scale is calibrated with the spin-average mass of 1S bottomonia at $T=0$, given by $\bar{M}_{1S}=(M_{\eta_b}+3M_{\Upsilon})/4$.}
	\label{fig:meff-zero-temperature}
\end{figure*}

Bottomonium correlators at zero temperature are presented in \autoref{fig:meff-zero-temperature}, shown in the form of effective masses, $M_{\text{eff}}$, defined as $aM_{\text{eff}}(\tau)=\log[C(\tau)/C(\tau+a)]$. $\Upsilon$ and $\chi_{b0}$ are taken as examples for S-wave and P-wave bottomonia respectively; other channels show very similar behavior. To remove the energy shift constant inherent in the NRQCD formalism, hereafter all vertical scales are calibrated with the spin-averaged mass of 1S bottomonia. For the 1S/1P states, comparisons between measurements from Gaussian and wave-function-optimized operators are given in \autoref{fig:meff-zero-temperature}, to check the difference from the choice of various extended operators. Deviations in the effective masses of $\Upsilon$(1S) and $\chi_{b0}$(1P) only appear at very small $\tau$, but are almost non-existent within the long plateau. This indicates
that the ground-state energy in a given channel is well-determined and does not depend
on the choice of the meson operators.

Both Gaussian and wave-function-optimized operators are effective in overlapping with targeted states, as clear plateaus can be found from about $\tau \simeq 0.3$ fm in \autoref{fig:meff-zero-temperature}. 
Compared to measurements with point-like operators, the use of extended operators leads to an earlier onset of plateaus at times much shorter than the inverse temperature $1/T$. This provides more data points for extracting in-medium bottomonium properties at finite temperatures, where the maximal temporal extent is limited by $1/T$.

We also fit the bottomonium correlators in vacuum with a single-exponential ansatz within the plateau region indicated in \autoref{fig:meff-zero-temperature}. The fit ranges,  $[\tau^{\text{vac}}_{\text{min}},\tau^{\text{vac}}_{\text{max}}]$, are carefully chosen to ensure the dominance of the ground state contribution, where $\tau^{\text{vac}}_{\text{max}}$ is the largest temporal separation $\tau$ at which $\sigma_{M_{\text{eff}}}(\tau)/M_{\text{eff}}(\tau) \leq  5\%$ is satisfied. The lower bound of the range, $\tau^{\text{vac}}_{\text{min}}$, is the first $\tau$ where the excited-state contribution to the effective mass, $\delta M(\tau)$, is less than the statistical uncertainty, $\sigma_{M_{\text{eff}}}(\tau)$, i.e., $\delta M(\tau) < 25\% \times \sigma_{M_{\text{eff}}}(\tau)$. The extracted energy levels of different bottomonium states are found to be consistent with the experimental results from PDG~\cite{ParticleDataGroup:2024cfk}, except for the hyperfine splitting of the 1S bottomonia, which is particularly sensitive to the short-distance physics.

\subsection{In-medium case}
\begin{figure*}[!htp]
	\centering
	\includegraphics[width=0.3\textwidth]{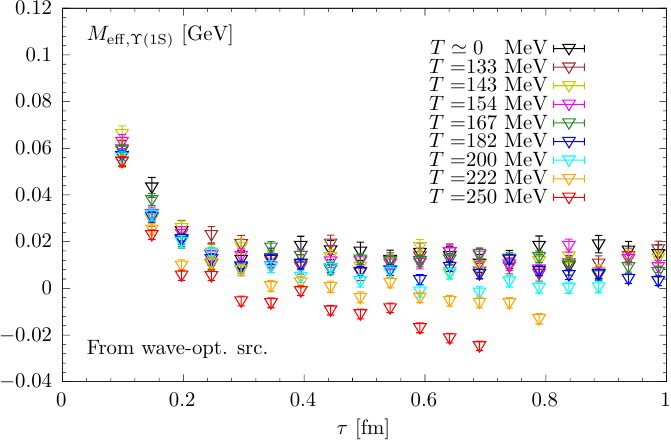}
	\includegraphics[width=0.293\textwidth]{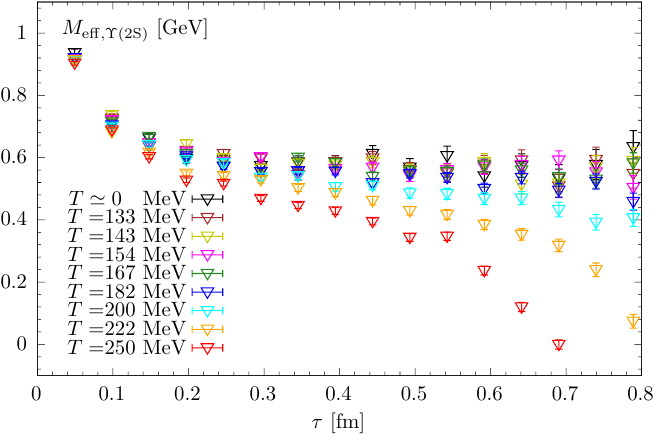}
	\includegraphics[width=0.3\textwidth]{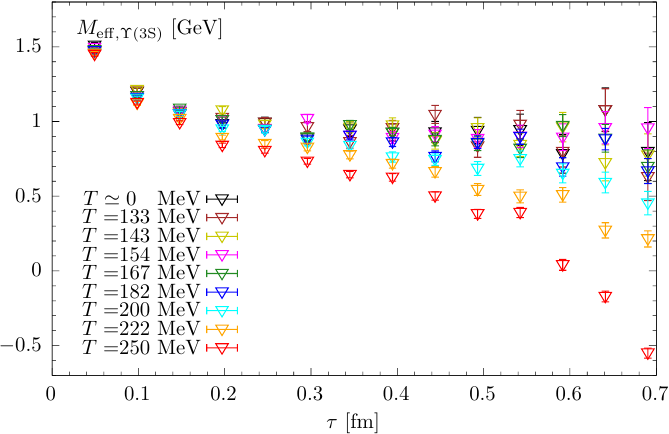}
    \includegraphics[width=0.295\textwidth]{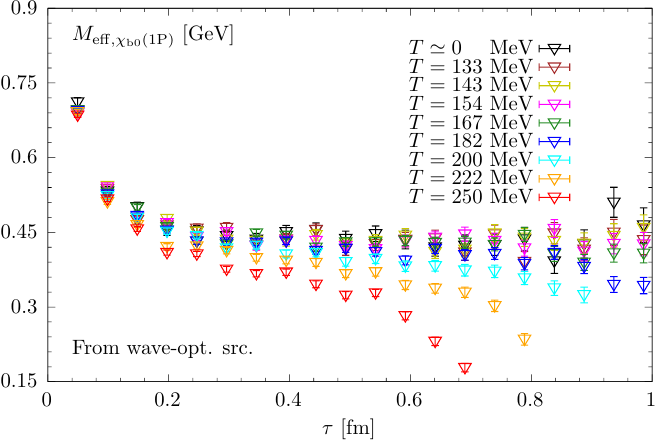}
    \includegraphics[width=0.3\textwidth]{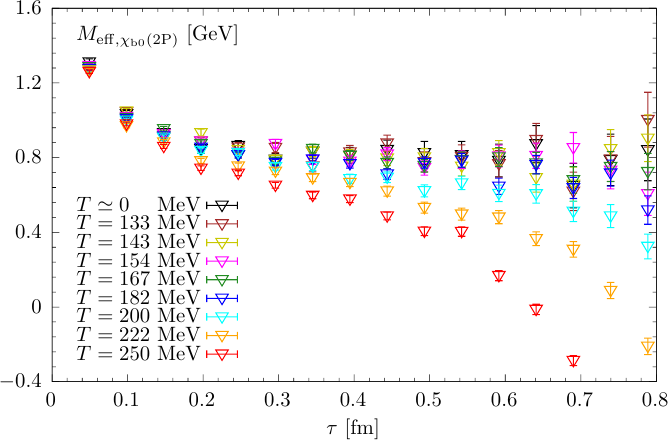}
    \includegraphics[width=0.3\textwidth]{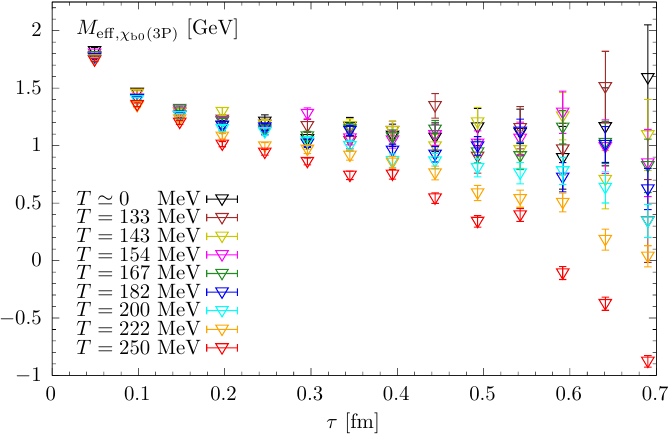}
  	\caption{Temperature dependence of effective masses for $\Upsilon(n$S) (top) and $\chi_{b0}(n$P) (bottom) with $n=$1, 2 and 3, measured using wave-function-optimized operators. The vertical scale is calibrated with the spin-average mass of 1S bottomonia at $T=0$.}
	\label{fig:meff-finite-temperature}
\end{figure*}

The temperature dependence of bottomonium correlators from wave-function-optimized operators at finite temperatures is shown in \autoref{fig:meff-finite-temperature} in terms of the effective masses, taking $\Upsilon$ and $\chi_{b0}$ as examples. For all states presented in \autoref{fig:meff-finite-temperature}, the effective masses exhibit mild temperature dependence within $\tau\lesssim 0.1$ fm for the whole temperature region under our consideration. 
Thermal effects become more significant as the temporal separation and temperature increase and are more pronounced for P-wave states compared to S-wave states. A plateau seems to exist until the temperature increases to 182 MeV for $\Upsilon$(1S) and 167 MeV for $\chi_{b0}$(1P), suggesting that these two ground S- and P-wave bottomonia can exist as bound states above the crossover temperature with roughly the same masses as in vacuum. At sufficiently high temperatures, the effective masses exhibit a sharp decrease around $\tau \simeq 1/T$. This drop becomes steeper with increasing temperature and is more pronounced for higher excited states.

\begin{figure*}[!htp]
	\centering
	\includegraphics[width=0.4\textwidth]{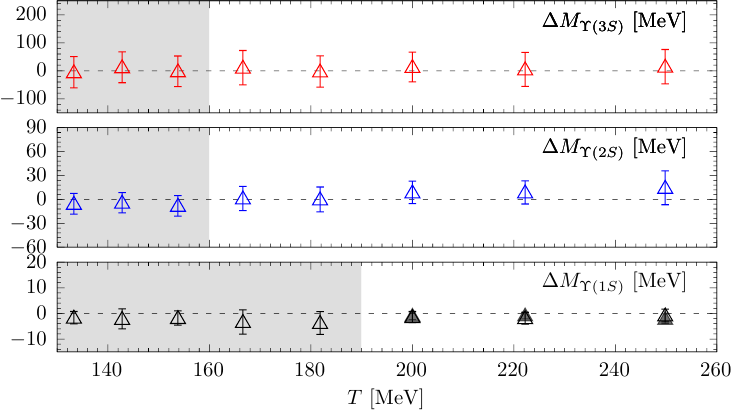}
	\includegraphics[width=0.4\textwidth]{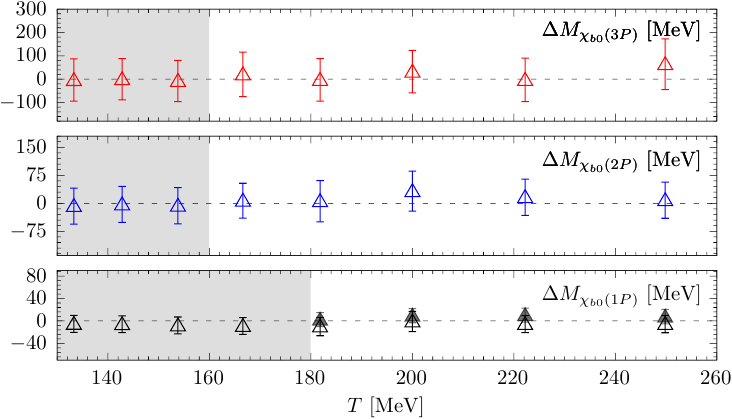}
	\\
	\includegraphics[width=0.4\textwidth]{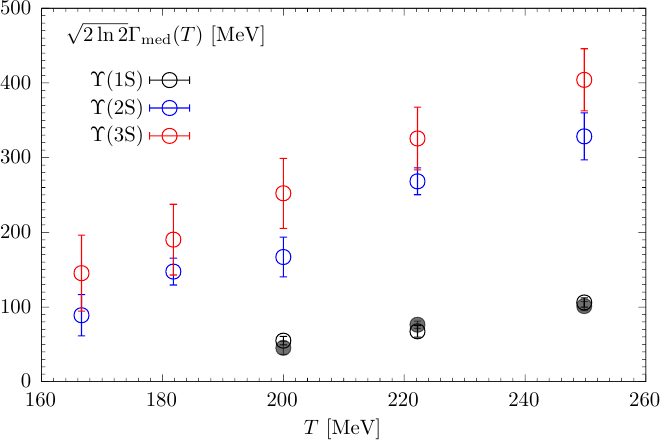}
	\includegraphics[width=0.4\textwidth]{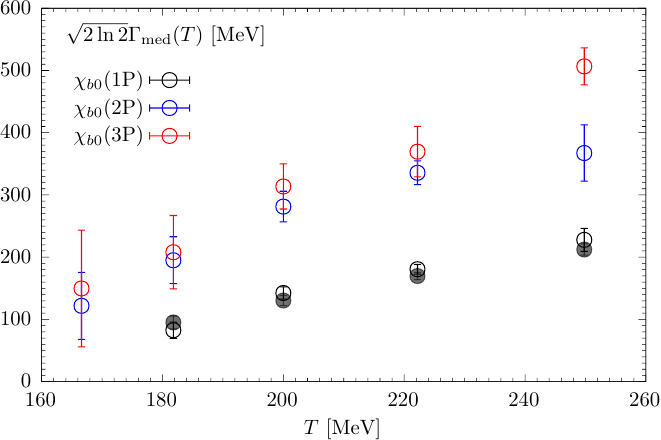}
	\caption{Temperature dependence of mass shifts (top) and in-medium widths (bottom), defined as the Gaussian widths at half maximum height, for $\Upsilon(n$S) and $\chi_{b0}(n$P) with $n=$1, 2 and 3 obtained from Gaussian fits on the continuum-subtracted correlators for the temperatures outside the gray-shaded area. Open points are from fits on the continuum-subtracted correlators calculated using wave-function-optimized operators, while filled points from Gaussian-smeared operators. For the temperatures located in the gray-shaded area,  a single-exponential fit is performed on the continuum-subtracted correlators.}
	\label{fig:inmedium-width-mass}
\end{figure*}

As for the effective masses calculated from Gaussian-smeared operators, qualitatively the same behavior can be seen, but there are quantitative difference compared with the results from wave-function-optimized operators. However, we will see that the in-medium properties from different extended operators are consistent with each other, as shown in \autoref{fig:inmedium-width-mass}.

To isolate the part that is more sensitive to the in-medium properties of bottomonia from the spectral function, a continuum-subtracted correlator can be defined~\cite{Larsen:2019bwy,Larsen:2019zqv} as follows. The spectral function is generally expected to contain several states below the open-bottom threshold, with numerous higher states forming a continuum at large frequencies. Thus, the spectral function can be 
decomposed into two parts, i.e., $\rho(\omega, T)=\rho_{\text{med}}(\omega, T)+\rho_{\text{cont}}(\omega)$. The continuum part, $\rho_{\text{cont}}$, is considered to be temperature independent, aligning with the temperature independent behavior of the effective masses at small $\tau$ in \autoref{fig:meff-finite-temperature}. The in-medium part of the spectral function, $\rho_{\text{med}}(\omega, T)$, at zero temperature can be parameterized as $\rho_{\text{med}}(\omega, T=0)=A\delta(\omega-M)$ with $M$ the mass of a certain state targeted for projection. This is because the extended operators used in this work have optimized overlap with such a target state and make the contribution from undesired states largely suppressed, as shown in \autoref{fig:meff-zero-temperature}. Given these discussions, $\rho_{\text{cont}}$ can be estimated at zero temperature through $\rho_{\text{cont}}=\rho(\omega, T=0)-A\delta(\omega-M)$, leading to
\begin{align}
	C_{\text{cont}}(\tau) = C(\tau, T=0) - A e^{-M\tau} \,,
    \label{eq:continuum-part-correlator}
\end{align}
where \autoref{eq:definition-nonrelativistic-spectra-function} was used, and the parameters $A$ and $M$ are extracted from fits to the exponential decay of vacuum correlators in the range $[\tau^{\text{vac}}_{\text{min}},\tau^{\text{vac}}_{\text{max}}]$ discussed in \autoref{subsec:vacuum}. Thus, the continuum-subtracted correlator is defined as~\cite{Larsen:2019bwy,Larsen:2019zqv}
\begin{align}
	C_{\text{sub}}(\tau, T)= C(\tau, T) - C_{\text{cont}}(\tau)\,.
    \label{eq:continuum-subtracted-correlator}
\end{align}
Then, a continuum-subtracted effective mass can conventionally be defined as
\begin{align}
	aM^{\text{sub}}_{\text{eff}}(\tau, T)=\log[C_{\text{sub}}(\tau, T)/C_{\text{sub}}(\tau+a, T)]\,.
    \label{eq:continuum-subtracted-effective-mass}
\end{align}

\begin{figure*}[!htp]
	\centering
	\includegraphics[width=0.3\textwidth]{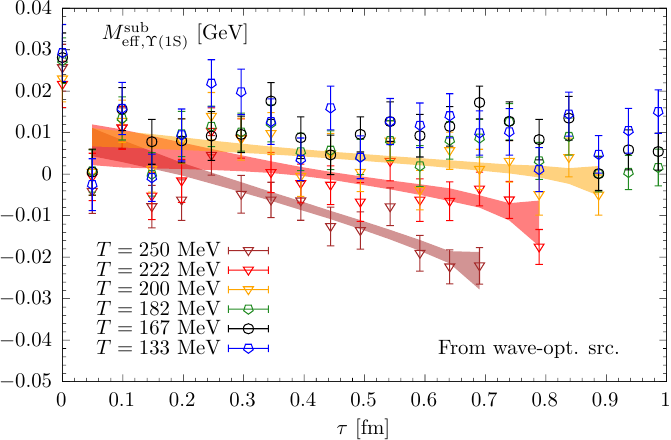}
    \includegraphics[width=0.3\textwidth]{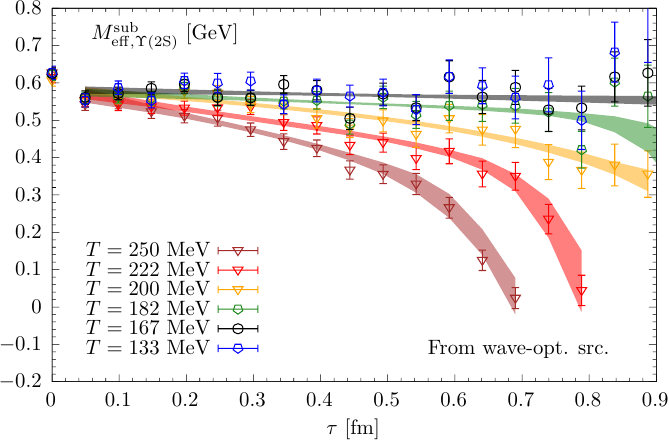}
    \includegraphics[width=0.3\textwidth]{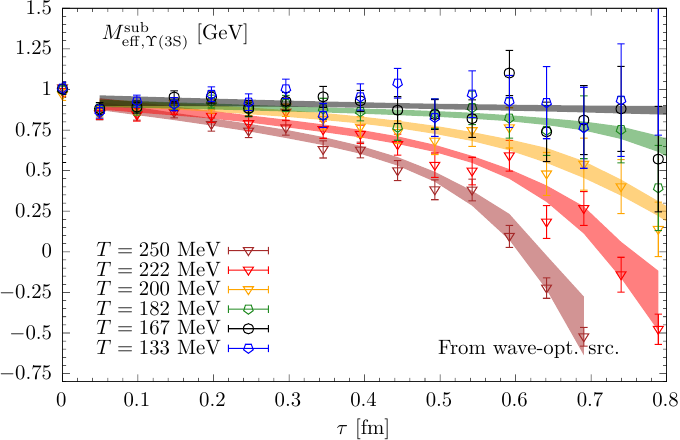}
	\includegraphics[width=0.3\textwidth]{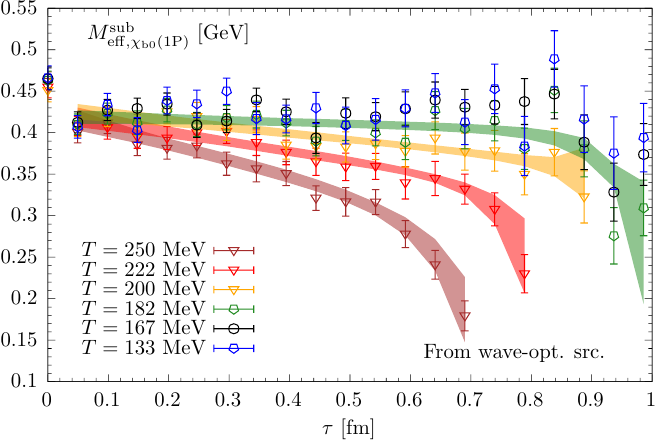}
  	\includegraphics[width=0.3\textwidth]{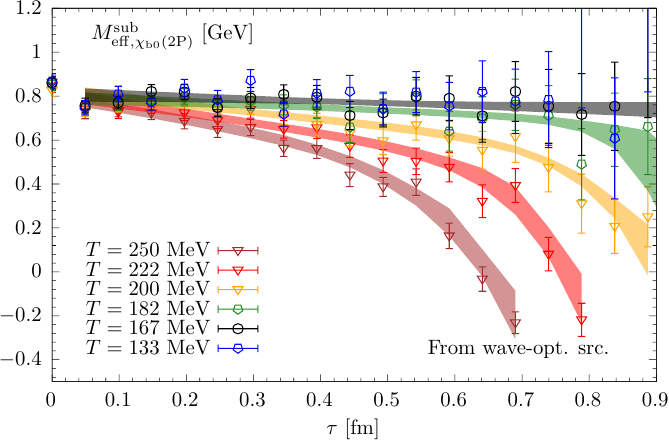}
  	\includegraphics[width=0.3\textwidth]{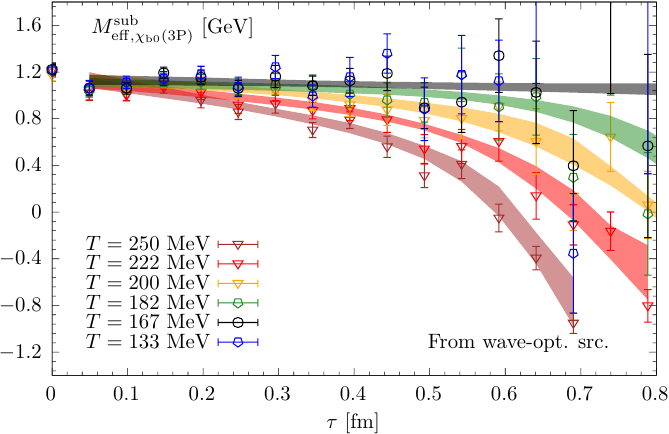}
    \caption{The continuum-subtracted effective masses with various temperatures for different excited states of $\Upsilon$ (top) and $\chi_{b0}$ (bottom). The bands are reconstructed from fits using the Gaussian ansatz (cf.~\autoref{eq:ansatz-inmedium-spectral-function}), as discussed in the text. The vertical scale is calibrated with the spin-averaged mass of 1S bottomonia at $T=0$.}
	\label{fig:subtracted-meff}
\end{figure*}

\autoref{fig:subtracted-meff} shows the temperature dependence of continuum-subtracted effective masses for $\Upsilon$ and $\chi_{b0}$ as an illustration, using wave-function-optimized operators. $M^{\text{sub}}_{\text{eff}}$ approaches the corresponding bottomonium mass in vacuum at small $\tau$, then shows a nearly linear decrease in the middle and finally ends with a sharp decline deviating from linearity around $\tau\sim 1/T$. The slope of $M^{\text{sub}}_{\text{eff}}$ in the middle $\tau$ range is considered to be closely related to the width of the bottomonium state, and grows larger for higher excited states and as the temperature increases. This is in line with the picture of sequential thermal broadening of bottomonia in the hot medium~\cite{Larsen:2019bwy,Larsen:2019zqv}. However, for some low temperatures, shown as $\Upsilon$(1S) at $T\lesssim$182 MeV and other states at $T\lesssim$167 MeV in \autoref{fig:subtracted-meff}, a roughly constant behavior with increasing $\tau$ can be observed, indicating that these states are almost not affected by the medium.

For the continuum-subtracted effective masses from Gaussian-smeared operators, similar behavior can be found, and the slope of the continuum-subtracted effective masses in the middle $\tau$ range is consistent
with that obtained from wave-function-optimized correlators. However, differences appear in the fast drop at the tail of $\tau$. 
This outcome is expected, as the portion of the spectral function significantly below the dominant peak influences the tail behavior of the continuum-subtracted effective masses~\cite{Bala:2021fkm}, which varies based on the choice of meson operators.

To extract in-medium properties, we adopt the Gaussian parameterization for the in-medium part of the spectral function, as proposed in Ref.~\cite{Larsen:2019bwy,Larsen:2019zqv}, i.e., 
\begin{align}
	\rho_{\mathrm{med}}(\omega, T) =
	 A_{\mathrm{med}}(T) \exp \left(-\frac{\left[\omega-M_{\mathrm{med}}(T)\right]^{2}}{2 \Gamma_{\mathrm{med}}^{2}(T)}\right)
	 +A_{\mathrm{cut}}(T) \delta(\omega-\omega_{\mathrm{cut}}(T))
	 \,.
    \label{eq:ansatz-inmedium-spectral-function}
\end{align}
Here, the $\delta$ term is effectively responsible for the part of the spectral function well below the main peak structure, reproducing the tail behavior of $M^{\text{sub}}_{\text{eff}}$ at large $\tau\sim 1/T$. The first term in \autoref{eq:ansatz-inmedium-spectral-function} leads to the linear decrease with respect to $\tau$ in $M^{\text{sub}}_{\text{eff}}$ within the middle $\tau$ region. The in-medium mass $M_{\mathrm{med}}(T)$ and width $\Gamma_{\mathrm{med}}(T)$ can be extracted by fitting the continuum-subtracted correlators using the ansatz \autoref{eq:ansatz-inmedium-spectral-function} with the help of \autoref{eq:definition-nonrelativistic-spectra-function}. The bands shown in \autoref{fig:subtracted-meff} are the continuum-subtracted effective masses reconstructed from the fit parameters in \autoref{eq:ansatz-inmedium-spectral-function} through \autoref{eq:definition-nonrelativistic-spectra-function}, \autoref{eq:continuum-subtracted-correlator} and \autoref{eq:continuum-subtracted-effective-mass}. These fits capture the behavior of $M^{\text{sub}}_{\text{eff}}$ well for $\Upsilon$(1S) and $\chi_{b0}$(1P) at $T\gtrsim$182 MeV and other excited states at $T\gtrsim$167 MeV. As discussed above, however, for the other cases at low temperatures, as indicated by the gray-shaded area in \autoref{fig:inmedium-width-mass}, the roughly constant behavior of $M^{\text{sub}}_{\text{eff}}$ suggests that a single-exponential ansatz is sufficient to describe the continuum-subtracted correlators at these temperatures. For the excited S- and P-wave states at $T=$167 MeV, the sudden drop-off in $M^{\text{sub}}_{\text{eff}}$ around $\tau \simeq 1/T$ is no longer visible. Thus, for these cases, the $\delta$ term corresponding to the tail of the spectral function can be ignored by setting $A_{\mathrm{cut}}(T)=0$ in the ansatz \autoref{eq:ansatz-inmedium-spectral-function} and omitting $2-4$ data points at the largest values of $\tau$ during the fits.

The temperature dependence of the in-medium mass $M_{\mathrm{med}}(T)$ and width $\Gamma_{\mathrm{med}}(T)$ extracted from the continuum-subtracted correlators, using the Gaussian parameterization for the in-medium spectra function, is presented in \autoref{fig:inmedium-width-mass}. The in-medium mass $M_{\mathrm{med}}(T)$ is shown in the form of a mass shift, i.e., the difference between $M_{\mathrm{med}}(T)$ and its vacuum counterpart, $\Delta M =M_{\mathrm{med}}(T)-M(T=0)$. The examples are provided for $\Upsilon$ and $\chi_{b0}$, and similar observations apply to other bottomonia. The overlaps of the filled and open black points in \autoref{fig:inmedium-width-mass} confirm that the in-medium properties do not depend on the choice of the extended operators.

The in-medium mass shifts $\Delta M$ in the top panel of \autoref{fig:inmedium-width-mass} are consistent with zero across all bottomonium states and the entire temperature range considered, with no clear temperature dependence observed. This indicates that the in-medium masses of various bottomonium states remain very close to their vacuum masses.  On the other hand, the in-medium widths, defined as the Gaussian widths at half maximum height, i.e., $\sqrt{2\ln 2} \Gamma_{\text{med}}$, in the bottom panel of \autoref{fig:inmedium-width-mass} show a clear increase with increasing temperature and are larger for higher excited states. A hierarchical pattern for the in-medium widths appears, corresponding to the hierarchy of bottomonium sizes: $\Gamma_{\text{med,1S(P)}}(T) < \Gamma_{\text{med,2S(P)}}(T) < \Gamma_{\text{med,3S(P)}}(T)$, while the difference between the widths of 2S(P) and 3S(P) states becomes less and less obvious as the temperature decreases.

There are other parametrizations for the in-medium spectral function that can reproduce the behavior of the continuum-subtracted correlators. In fact, the Gaussian form of the in-medium spectral function is not physically motivated, and the interpretation of Gaussian widths as the widths of bottomonium states is not well-defined. Assuming that the detailed shape of the spectral function away from the main peak structure is unimportant, the Gaussian width at half maximum height can be interpreted as the in-medium width of a bottomonium state, which is equal to $\sqrt{2\ln2}\Gamma_{\text{med}}$ and has been shown in \autoref{fig:inmedium-width-mass}. A more natural choice of the parametrization for the in-medium spectral function would be a Breit-Wigner (Lorentzian) form, and our further work~\cite{Ding:2025fvo} focuses on such a choice and comparisons between different parametrizations. However, the in-medium parameters extracted from the Gaussian parametrization could provide qualitative results and possibly  upper bounds for the in-medium width.

\section{Conclusion \label{sec:conclusion}}
We study the in-medium properties up to 3S and 3P excited bottomonia from lattice NRQCD calculations with two types of extended operators, for temperatures ranging from 133 to 250 MeV. From the temperature dependence of the bottomonium correlators, we find that all bottomonium states below the open-bottom threshold can exist as well-defined quasi-states above the crossover temperature, including 3P states. After extracting the in-medium widths and masses by using the Gaussian parametrization for the in-medium spectral function, we find that the bottomonium masses do not change compared to their vacuum values across the temperature range considered, while nonzero widths are observed for the different bottomonium states. These observations suggest that screening may not be the likely source of bottomonium dissociation in the medium. Additionally, we confirm that the in-medium properties of bottomonia are not affected by the choice of Gaussian-smeared or wave-function-optimized operators used in the measurements.

\acknowledgments
This work is supported partly by the National Key Research and Development Program of China under Contract No. 2022YFA1604900; the National Natural Science Foundation of China under Grants No.~12293064, No.~12293060 and No.~12325508, and by the U.S. Department of Energy, Office of Science, Office of Nuclear Physics through Contract No.~DE-SC0012704, through Heavy Flavor Theory for QCD Matter (HEFTY) topical collaboration in Nuclear Theory,  and within the framework of Scientific Discovery through Advance Computing (SciDAC) award Fundamental Nuclear Physics at the Exascale and Beyond. S. Meinel acknowledges support by the U.S. Department of Energy, Office of Science, Office of High Energy Physics under Award Number DE-SC0009913. W.-P. Huang acknowledges support from the Young Elite Scientists Sponsorship Program (Special Support for Doctoral Students) by CAST.

This research used awards of computer time provided by the National Energy Research Scientific Computing Center (NERSC), a U.S. Department of Energy Office of Science User Facility located at Lawrence Berkeley National Laboratory, operated under Contract No. DE-AC02- 05CH11231, 
on Frontier supercomputer in Oak Ridge Leadership Class Facility through ALCC Award 
,
and the PRACE awards on JUWELS at GCS@FZJ, Germany and Marconi100 at CINECA, Italy. Computations for this work were carried out in part on facilities of the USQCD Collaboration, which are funded by the Office of Science of the U.S. Department of Energy.  Some of the calculations were performed on the GPU cluster in the Nuclear Science Computing Center at CCNU (NSC$^3$) and Wuhan supercomputing center.

\bibliographystyle{JHEP.bst}
\bibliography{ref.bib}

\end{document}